\documentclass[twocolumn,showpacs,amsmath,amssymb,prb]{revtex4}

\usepackage{graphicx,rotating}
\usepackage{dcolumn}
\usepackage{bm}
\usepackage{epsfig}
\usepackage{caption}

\begin{document}

\title{Separate-Path Electron and Hole Transport
Across $\pi$-Stacked Ferroelectrics\\ for Photovoltaic Applications.}

\author{Ma\l{}gorzata Wawrzyniak-Adamczewska}
\affiliation{%
Faculty of Physics, A. Mickiewicz University, ul. Umultowska 85, 61-614
Pozna\'n, Poland
}%

\author{Ma\l{}gorzata~Wierzbowska}
\affiliation{%
Institut of Physics, Polish Academy of Sciences (PAS),
Al. Lotnik\'ow 32/46, 02-668 Warszawa, Poland
}%

\date{\today}

\begin{abstract}
Abstract: Electron and hole separate-path transport is theoretically found in 
the $\pi$-stacked organic layers and columns. This effect might be a solution 
for the charge recombination problem. 
The building molecules, named 1,3,5-tricyano-2,4,6-tricarboxy-benzene, 
contain the mesogenic flat aromatic part and the terminal dipole groups which make the
system ferroelectric. The diffusion path of the electrons cuts 
through the aromatic rings, while holes hop between the dipole groups. 
The transmission function and the charge mobilities, especially for the holes, 
are very sensitive to the distance between the molecular rings,
due to the overlap of the $\pi$-type orbitals. 
We verified that the separation of the diffusion paths 
is not destroyed by the application of the graphene leads.
These features make the system suitable for the efficient solar cells, 
with the carrier mobilities higher than these in 
the organometal halide perovskites.
\end{abstract}

\keywords{pi-stacking, ferroelectrics, photovoltaics, diffusion path,  
electronic transport, graphene electrodes}
\maketitle

\section{Introduction} 

\begin{figure*}
\vspace{5mm}
\centerline{
\includegraphics[scale=0.24]{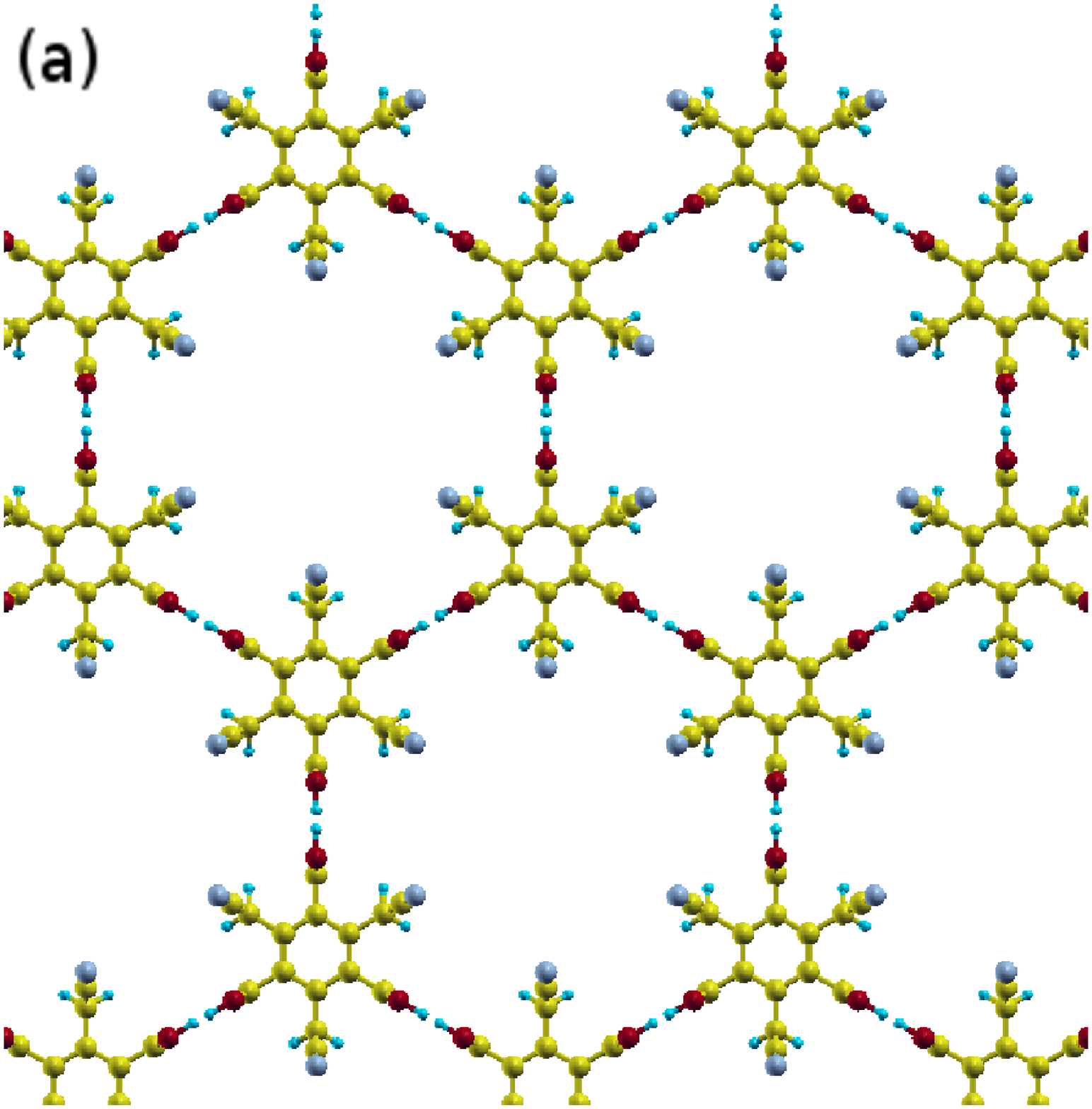}
\includegraphics[scale=0.24]{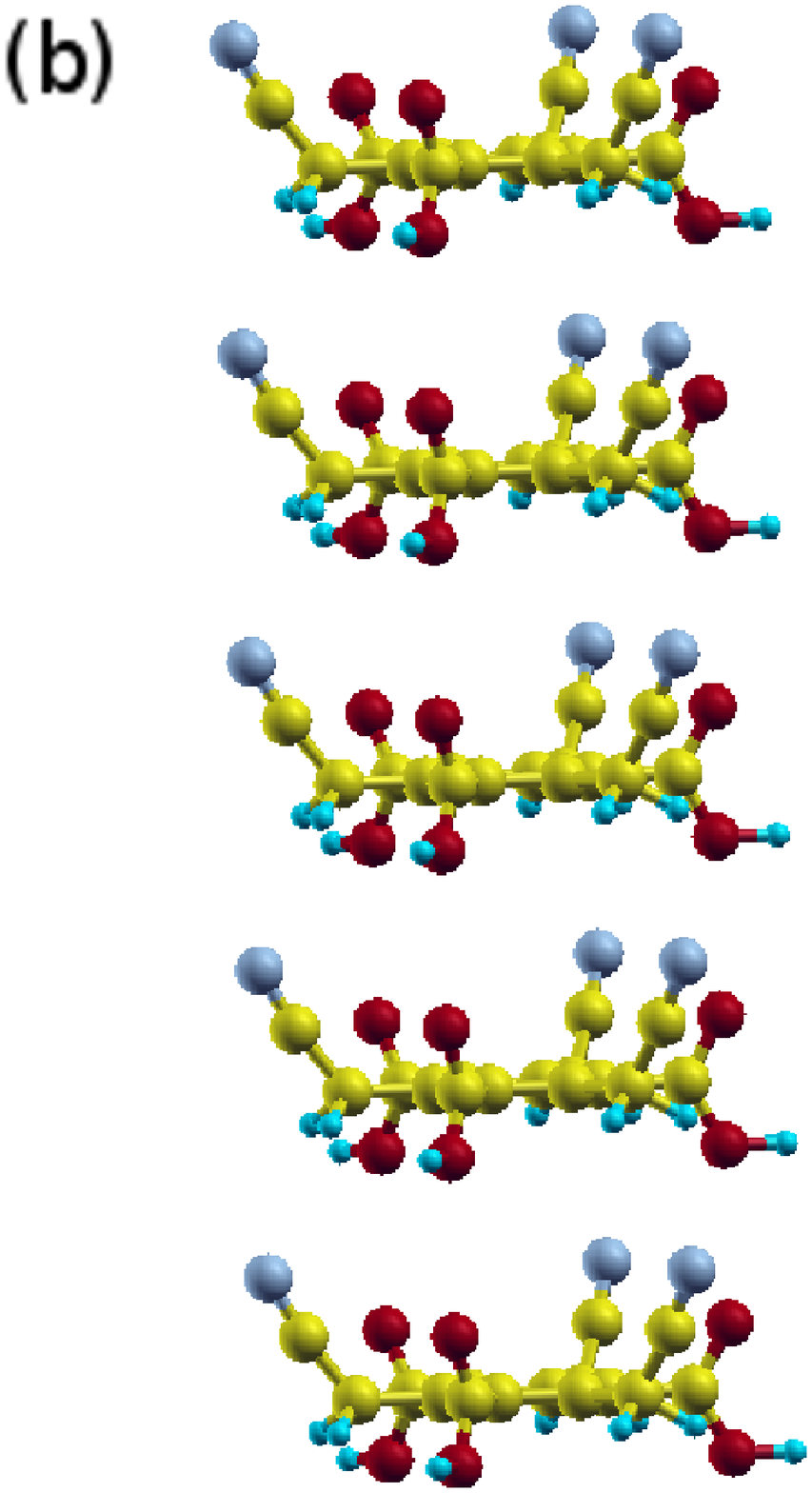}
\includegraphics[scale=0.2]{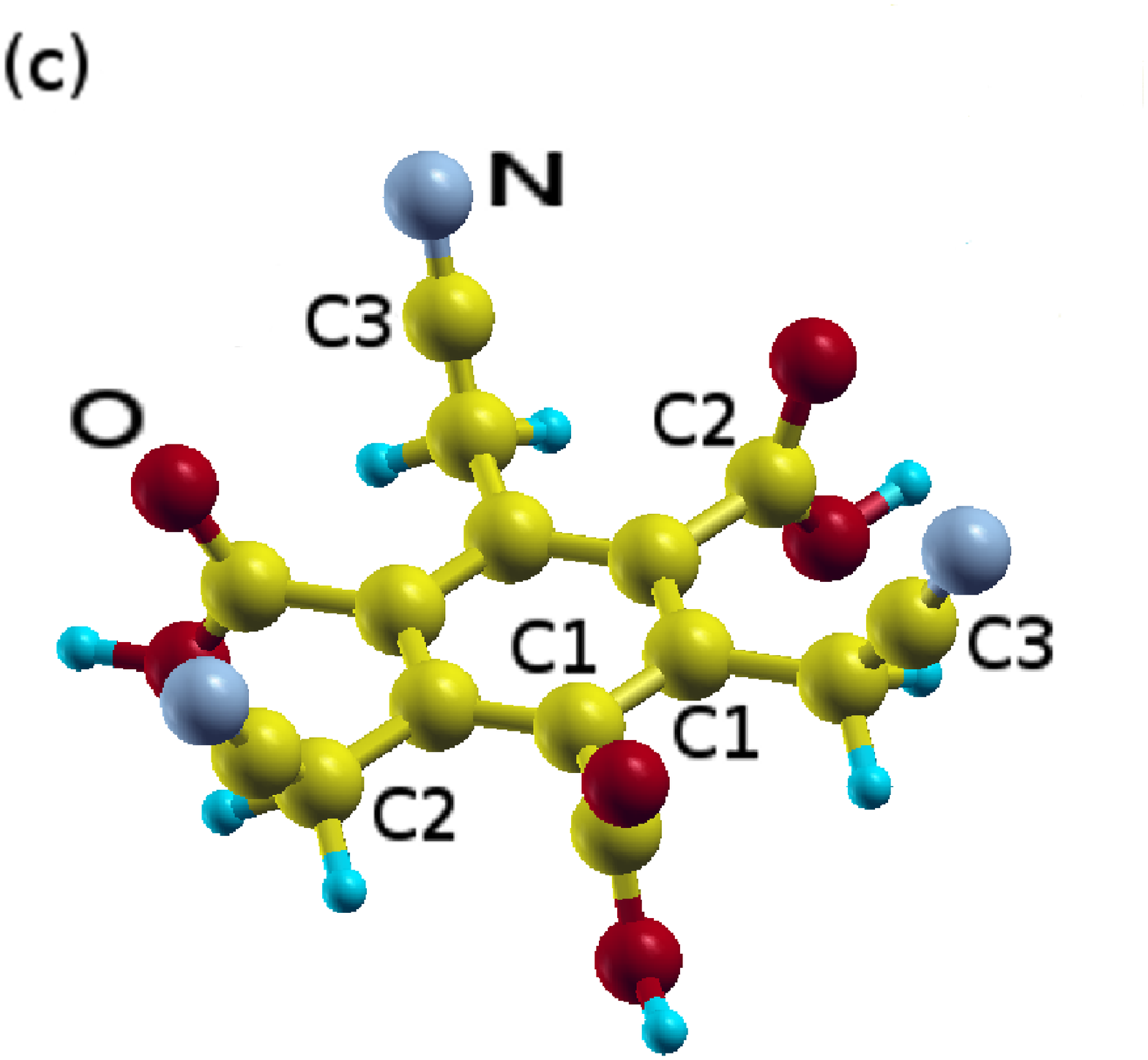}}
\caption{Atomic structure of investigated systems: (a) top view at molecular layers, (b) side
view at columnar stacking, (c) single molecule with the indexed atoms.}
\label{b1}
\end{figure*}
 
The effect of charge carriers recombination is a very 
unwanted phenomenon in the photovoltaic materials. There are various attempts 
to avoid this process in novel solar cell devices. One of the most popular
solution, valid for electrons as well as holes, is an addition of the charge 
trapping layers. These layers selectively permit carriers transport across 
the multi--junctions. For example, the polymer solar cells equipped with the 
PEDOT:PSS layers allow to extract hole carriers before they arrive at the 
anode.\cite{pedot:pss} Recently, similar hole trapping function has been 
found for the graphene oxide nanoribbons.\cite{GON} It has been shown, that 
the parameters of the metal--oxide--free methylammonium lead iodide 
perovskite--based solar cells are also improved by the charge transport 
layers.\cite{Olga}

On the other hand, the simultaneous trapping of the electrons and 
holes limits the photovoltaic efficiency. This is an inevitable effect in many
systems. The organometal halide perovskites, where the liquid ions\cite{cpl1} 
and surface states\cite{jpcl2} grab both types of carriers are the examples. 
Another performance key factor is the material structure,\cite{natnano1} 
particularly important in the solution processed organometal trihalide 
perovskite solar cells. For this kind of systems, planar heterojunctions are 
less effective than the mesosuperstructured perovskites, regarding the 
occurrence of the sub gap states, but superior in terms of carrier 
mobility.\cite{acsnano1,pccp1} Luckily, the efficiency grows again at very 
high fluences in the presence of the sub gap states,\cite{model1} and the 
trap-states density can be optimized in terms of the grain size of 
nanocrystals.\cite{science1} In addition, under temperature gradient, several 
structural phase transitions may occur in these materials, and this fact also 
influences the diffusion lengths.\cite{AFM1}

Modern approach to increase the solar cell efficiency goes towards 
the so--called bulk heterojunctions, where the granular structure of 
two materials, the donor-type (e.g. polymers) and acceptor-type 
(e.g. fullerenes), might be regular or graded.\cite{bulkhetero}   
It has been found that a larger grade of the composition corresponds to
wider charge transportation path and 
larger effective conductivity.\cite{bulkhetero}
Alternatively to the bulk heterojunctions, the ordered
donor-acceptor materials -- integrated heterojunctions --
have been synthesized.\cite{IHJ} They are composed of two molecules
integrated in the covalent organic frameworks. Recently, 
these kind of materials gain an attention for their photo-active 
properties.\cite{IHJ,Bein-3,Bein-4,Bein-5}  

An addition of the inter-layer aligns the energy levels of 
the multijunction device, in such a way that they form a cascade, efficiently 
reduces the electron-hole recombination.\cite{cascade} 
The above effect is intensified if one uses the 
high/low work-function material for the anode/cathode. This tuning might 
be done by an adsorption of the dipole moieties.\cite{workOH}

In this work, we present a new solution to the charge-carriers recombination 
problem -- the electron and hole separate-path transport. 
The considered system has been described in our previous work\cite{previous} 
for its ferroelectric properties. 
The studied layers are characterized by the cascade energy 
levels alignment, gradual donor--acceptor layers sequence, and the 
dipole-proximity tunable work function of the graphene electrodes. 
In Figs. 1(a)-(c), 
we present the system which consists of the flat benzene-based molecules, 
terminated with the cyano groups (NCCH$_2$), that are responsible for the 
ferroelectric properties, and the carboxy groups (OCOH), that form the 
hydrogen bonds within the planes. The molecules are named 
1,3,5-tricyano-2,4,6-tricarboxy-benzene and their chemical formula reads
C$_6$-3(NCCH$_2$)-3(OCOH). The band gap of the 2D layer is around 3.6 eV, 
while for the columnar stacks is 2.95 eV. We focused on localization of the 
diffusion paths, but also analyzed the electron and hole mobilities within 
the planes and across the stacks.

The considered building molecules represent the most simplified 
model molecules possessing the desired properties. Systems similar to 
the considered ones, could be realized in industry by modifications of the 
existing columnar liquid crystals,\cite{LC} or {\it via} the self-assembly of 
the 2D molecular systems, for example on graphene. This has been reported in  
the experimental\cite{graphorder} and theoretical works.\cite{Lublin} 
Alternatively, one could base on the production processes of
the covalent organic frameworks\cite{Bein-1,Berkeley} using the external
electric field in order to orient all dipole groups ferroelectrically,
and avoid formation of the antiferroelectrically coupled domains.\cite{Horiuchi}

The ferroelectric hydrogen--bonded organic frameworks\cite{previous,Sobol} 
could be obtained by modifications of the covalent integrated heterojunctions\cite{IHJ} or 
bulk organic ferrolectrics.\cite{Horiuchi-2,Horiuchi-3}
We expect that these frameworks would be interesting for the solar cell
community, opening a new door for the efficient photovoltaic devices. 

\begin{figure}
\includegraphics[scale=0.3]{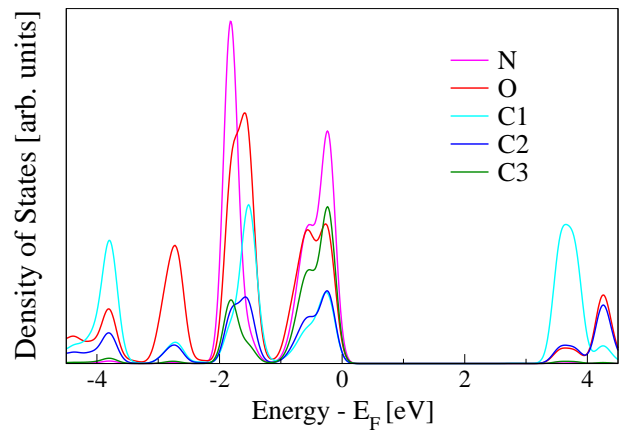}
\caption{DOS projected at groups of atoms indexed in Fig. 1(c),
for the top layer of the six monolayers system in the AA-type stacking.}
\label{b2}
\end{figure}

\begin{figure*}
\centerline{
\includegraphics[scale=0.20]{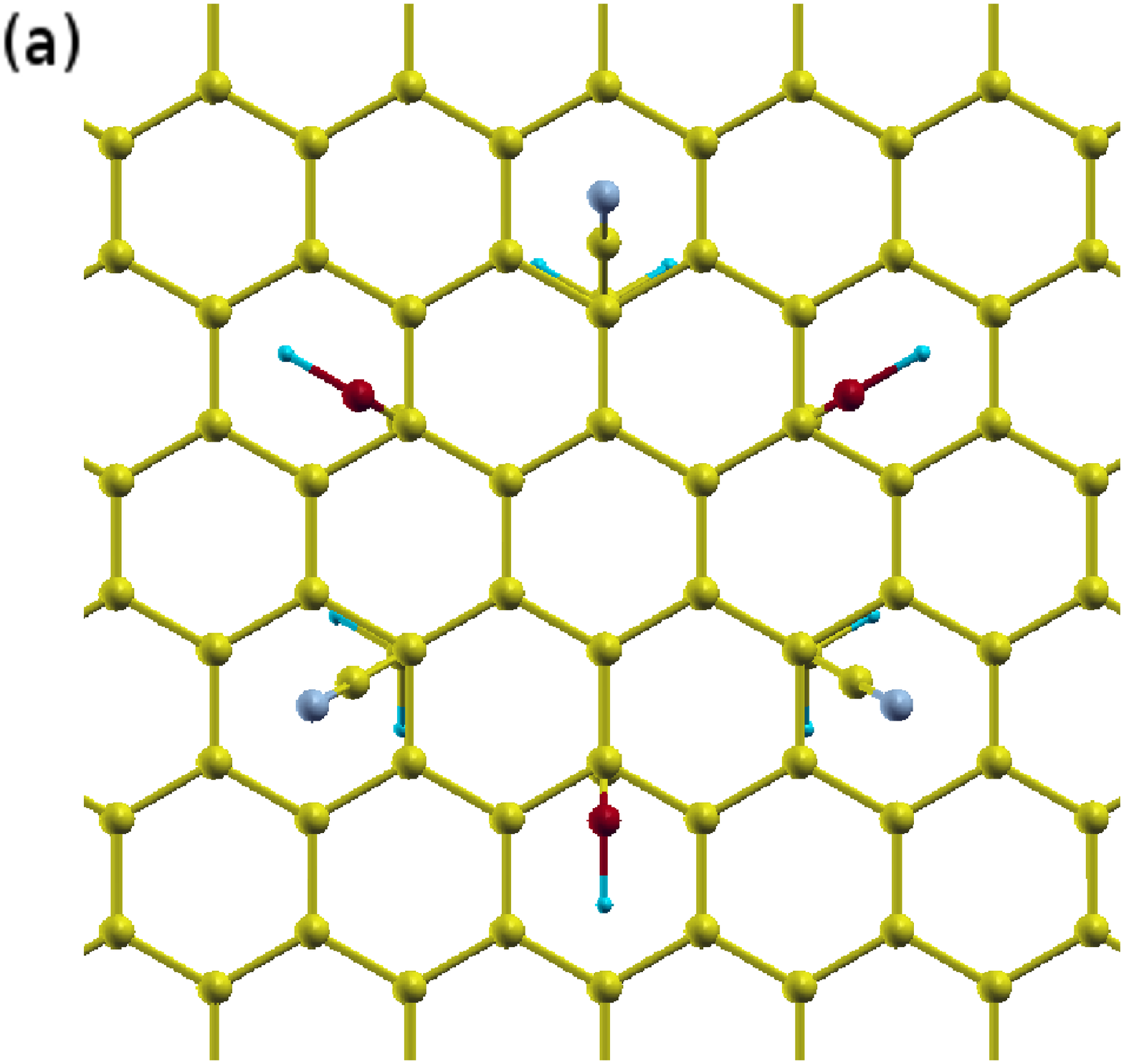}  \hspace{0.2cm}
\includegraphics[scale=0.2]{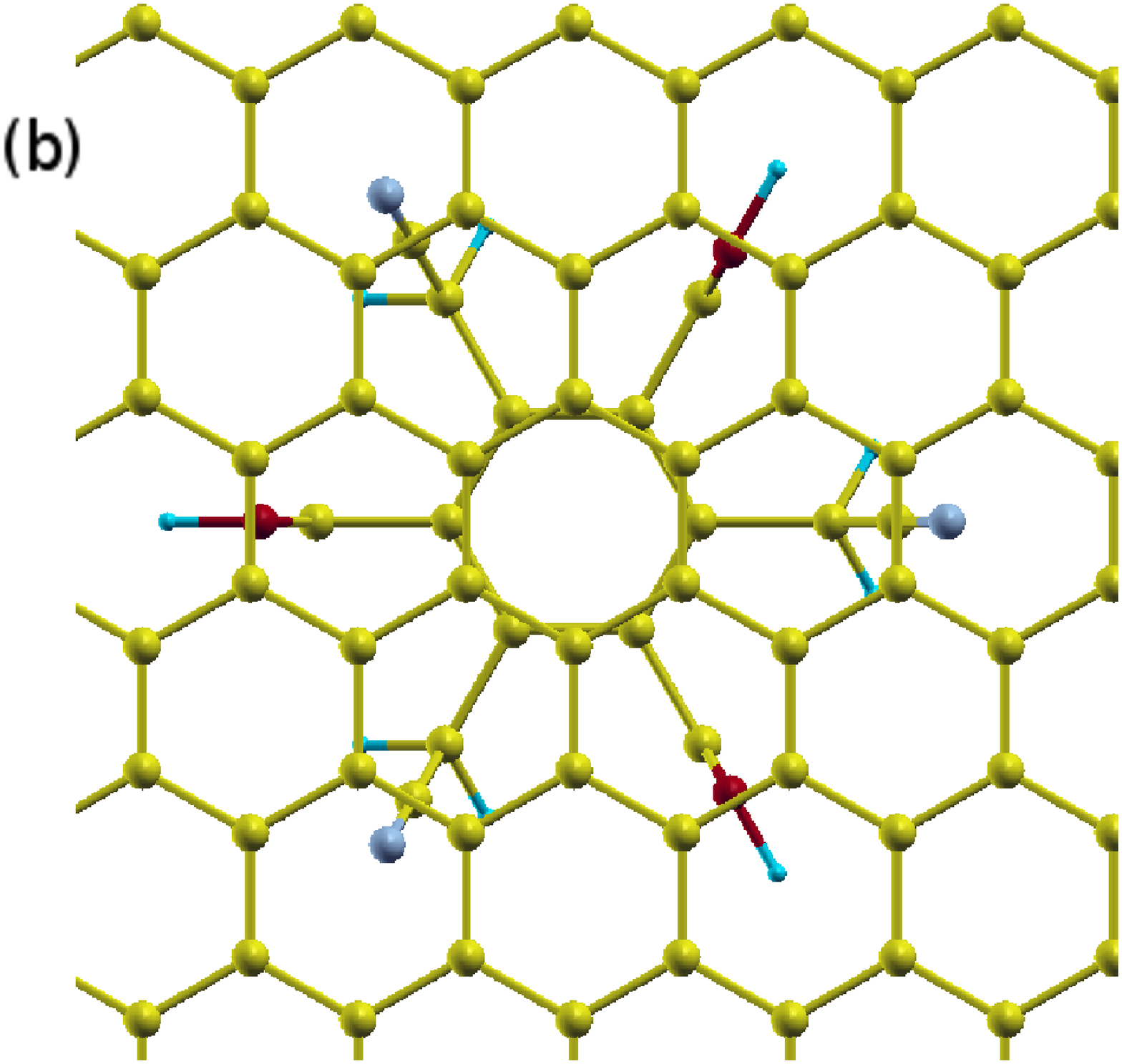}  \hspace{0.2cm}
\includegraphics[scale=0.2]{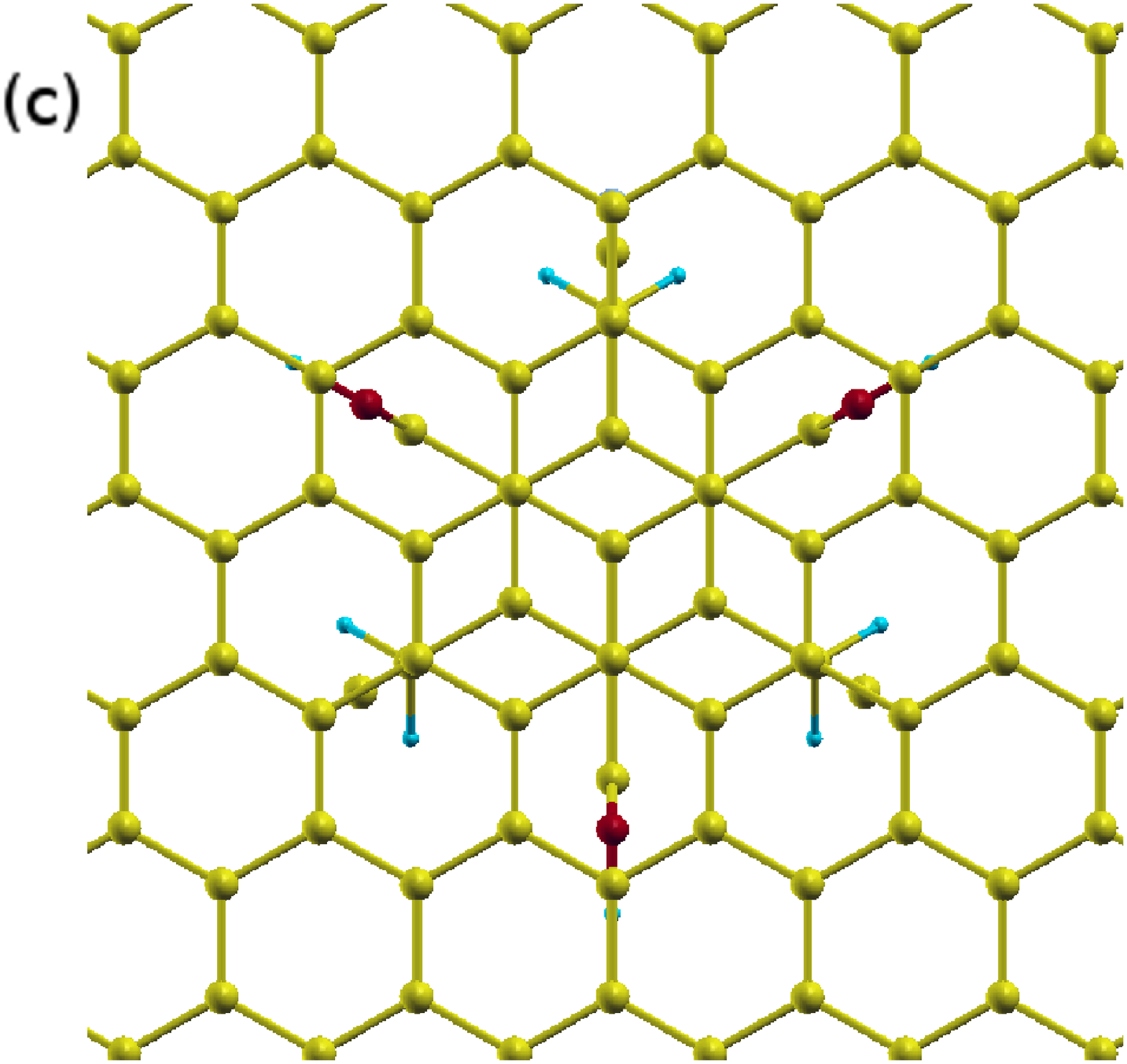}} \vspace{0.2cm}
\centerline{
\includegraphics[scale=0.22]{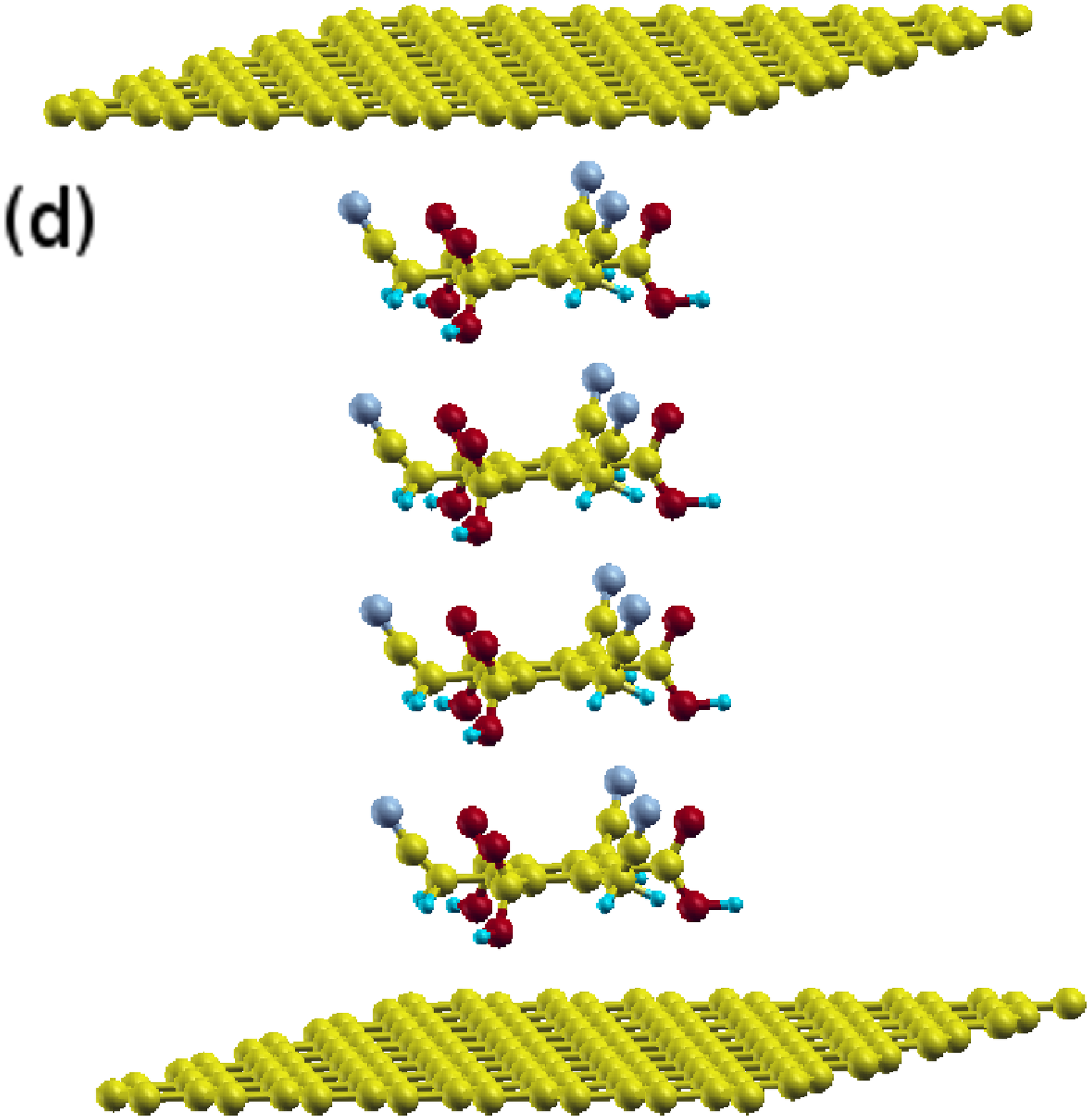}
\includegraphics[scale=0.25]{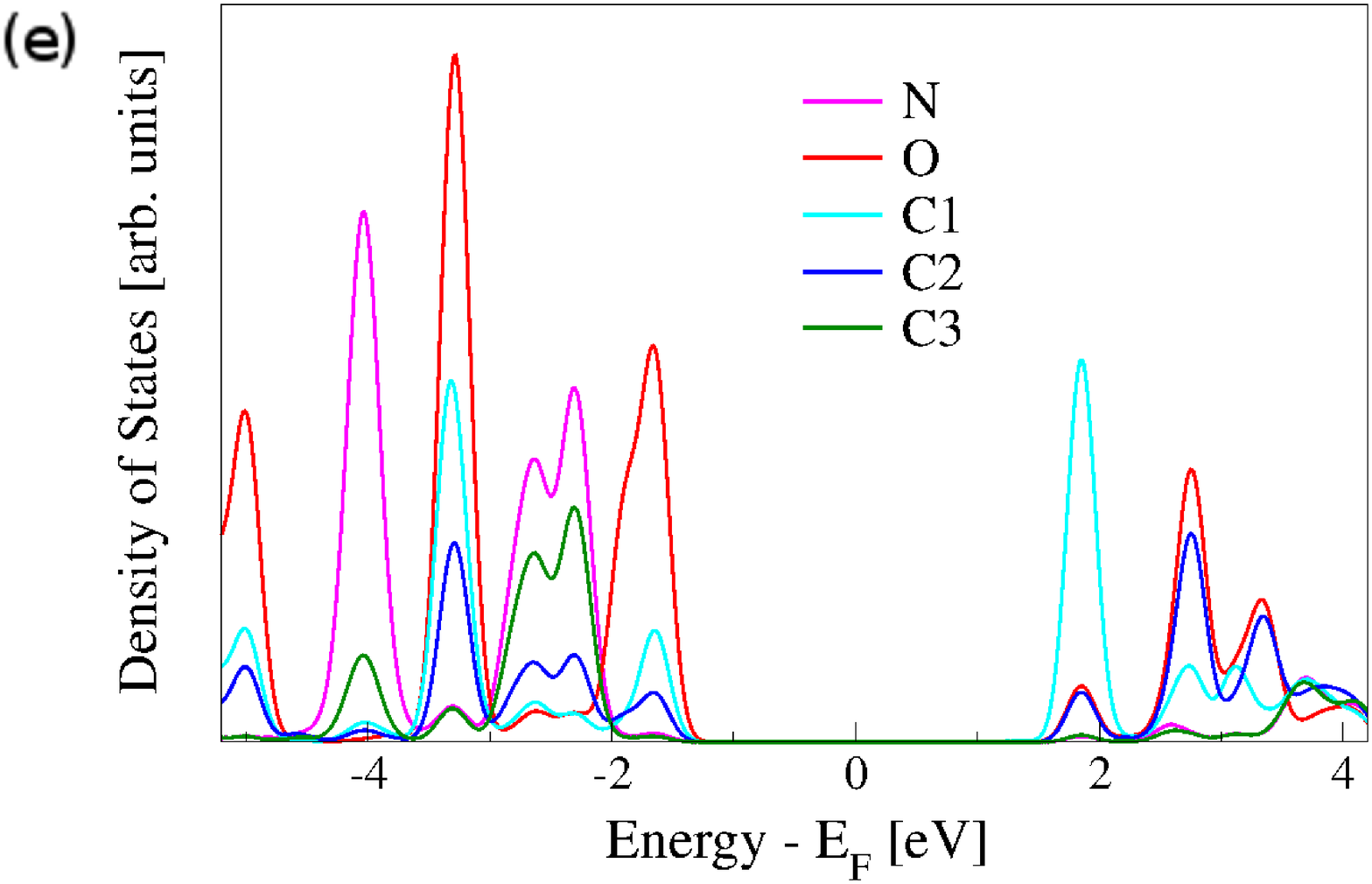}}
\caption{A column of four molecules between the graphene leads:
(a) top view at the AA-type stacked system,
(b) R30$^\circ$-stacked system,
(c) AB-stacked case,
(d) side view,
(e) DOS for the top molecule of the AA-stacked system,
projected at the groups of atoms indexed in Fig. 1(c).}
\label{b3}
\end{figure*}

\section{Method} 

We calculated various material properties derived from 
the electronic states obtained within the density functional theory (DFT).     
All calculations, except for the transmission function, were performed 
employing the {\sc Quantum ESPRESSO} suite of codes.\cite{qe}
The density of states (DOS), band structures, and carrier mobilities for 
the one--, two-- and three--dimensional structures were obtained on top of 
the Bloch states, represented in the plane-wave basis set and the 
pseudopotentials for the core electrons.
For most of the calculations, if not stated explicitly, the 
exchange--correlation functional was chosen for the gradient corrected 
Perdew-Burk-Erzenhof parametrization.\cite{PBE} 

In order to accurately interpolate the band structures, we used the wannier90 
package,\cite{w90} which enables finding the maximally-localized 
Wannier functions for the composite bands.\cite{wan,RMP} Next, 
we projected the band structures on the optimized Wannier functions, which 
were centered around chosen atoms.

The mobilities $\mu$ are defined as the ratio of the conductivity $\sigma$ 
and the carrier density $\rho$, that is obtained from the 
quadrature of the DOS, and the system volume V, $\mu=\sigma/(e\rho)$,
whereas $e$ denotes the electron charge. The conductivity 
is defined as follows:
\begin{eqnarray}
\sigma_{ij}(E) & = & \frac{1}{V} \; \sum_{n,{\bf k}} \; v_i(n,{\bf k}) v_j(n,{\bf k})
\; \tau_{n{\bf k}} \; \delta (E-E_{n,{\bf k}}),  \nonumber \\
v_i(n,{\bf k}) & = & \frac{1}{\hbar} \; \frac{\partial E_{n,{\bf k}}}{\partial k_i},
\nonumber
\end{eqnarray}
$v_i(n,{\bf k})$ is the band velocity, $E_{n,{\bf k}}$ is the band dispersion,
and $\tau_{n{\bf k}}$ denotes the relaxation time dependent on 
the band index $n$ and the reciprocal space ${\bf k}$. We used
the constant $\tau$ approximation within this work, with its value set to 10 fs.
The first derivatives of the energy bands in the Brillouin zone were 
obtained using the BoltzWann post-processing code\cite{boltz} from the 
wannier90 package.

The total relaxation time is a sum of many contributions: elastic (due to acoustic phonons),
nonelastic (due to the optical phonons), and the ionic impurities effects.
\begin{eqnarray}
\frac{1}{\tau} & = & \frac{1}{\tau^{ac}} + \frac{1}{\tau^{opt}} + \frac{1}{\tau^{imp}},
\nonumber 
\end{eqnarray}
The elastic scattering theory and  the Fermi golden rule\cite{bardeen} reads
\begin{eqnarray}
\frac{1}{\tau^{ac}_{\bf k}} & = & \sum_{\bf k'} \frac{2\pi}{\hbar} \; \delta
( \epsilon_{\bf k}-\epsilon_{\bf k'}) \; |M_{{\bf k},{\bf k'}}|^{2} \;
(1-cos(\theta )) \nonumber \\
&& |M^{ac}_{{\bf k},{\bf k'}}|^{2} \; = \; \frac{k_B T \; E_1^2}{C_{ii}}. \nonumber
\end{eqnarray}
The elastic constant, $C_{ii}$, can be obtained from
the total energies, $E$ and $E_0$, of the strained and equilibrium system
with respect to the dilation ($\Delta l/l_0$), following the formula
$(E-E_0)/V_0 = C_{ii}(\Delta l/l_0)^{2}/2$. The value $E_1$ can be obtained
from the valence and conduction band energy change
$\Delta \epsilon_{VBT(CBM)}$ \-- for the holes and electrons, respectively \--
with respect to the dilation ($\Delta l/l_0$). Theta denotes the angle 
between the vectors {\bf k} and {\bf k'}, and for the 1D structure can be equal only 
0 or 180 deg. Contribution to the relaxation time is nonzero only
for the backward scattering.

The nonelastic scattering due to the electron-phonon coupling with 
the optical phonons is much lower than that from the acoustic phonons. 
This fact is highlighted in the work about the naphthalene crystal.\cite{elast,napht} 
Importance of  the acoustic phonons has been reported also for the organic perovskites.\cite{srep}
For the ordered quasi-1D systems, in absence of the elastic and ionic contributions, 
purely electron-phonon scattering might lead to the mobilities, which are three orders 
of magniture larger than these measured for the 3D-disordered crystals of the 
same molecules.\cite{Casian}
    
In contrast to the nonelastic effects, the scattering in the presence of 
the charge impurities often predominates in the organic crystals.\cite{srep,elast,penta}   
In this case, the screened Coulomb potential leads 
to the scattering matrix elements\cite{ion}
\begin{eqnarray}
|M^{imp}_{{\bf k},{\bf k'}}|^{2} & = & 
\frac{n\;Z_{ion}^2\;e^4}{V_0^2\;(\varepsilon_r\varepsilon_0)^2\;
(L_D^2+|{\bf k'}-{\bf k}|^2)^{-2}} \nonumber \\
&& L_D = \sqrt{\varepsilon_r\varepsilon_0k_BTN_0\;/\;e^2}. \nonumber
\end{eqnarray}
where $n$ is the concentration of the ionic impurities, $Z_{ion}$ is the charge
of the impurity, $L_D$ is the Debye screening length with the free charge concentration
$N_0$, the relative permitivity of the material $\varepsilon_r$ and the dielectric
constant of the vacuum $\varepsilon_0$. The dielectric constant of the 1D molecular
systems was obtained using the Quantum Espresso postprocessing tool epsilon.x.\cite{eps} 

The transmission functions were calculated using the DFT method, as implemented 
in the {\sc siesta}-3.0 package.\cite{siesta1,siesta2} The software is based on the 
pseudopotential approach and uses a finite range localized basis sets. The 
transport computations\cite{Brandbyge02} use the non--equilibrium Green's 
functions, represented in terms of the solutions of the Kohn--Sham 
Hamiltonian.\cite{KS1,KS2} Both the local-density approximation (LDA) for the 
electron exchange and correlation in the Ceperley-Alder 
parametrization,\cite{CA} as well as the PBE parametrization, were used 
to compare the transmission functions for the system geometries optimized 
in each method. 

The system geometry, used for the transport calculations, has been 
arranged as follows: the molecule was sandwiched between the 
two graphene planes (upper and lower). The upper and lower graphene sheets 
were oriented in the AA-type stacking with each other and with the molecular 
aromatic-ring. In order to provide the charge flow, for biased system, from 
the upper graphene sheet across the molecule to the lower graphene sheet 
(or vice versa), both graphene sheets were terminated. The edge atoms of 
graphene were additionally passivated with the hydrogen atoms. The extensions 
of the upper and lower graphene planes formed the left and right electrodes. 
Thus, the scattering region consists of the molecule, as well as the
upper and lower discontinuous graphene sheets. The supercell contains 850 
atoms. The interplanar graphene distance in the geometry optimized with 
the LDA scheme is 8.6 $\AA$, 
and in the GGA-structure, it is 9.8 $\AA$. 
The maximal examined value of the $k$--sampling of the Brillouin zone, along 
the passivated graphene--edge direction, was $k=10$. Further details are 
included in the supporting information.

\begin{figure*}
\vspace{4.5cm}  \centerline{
\includegraphics[scale=0.25,angle=-90.0]{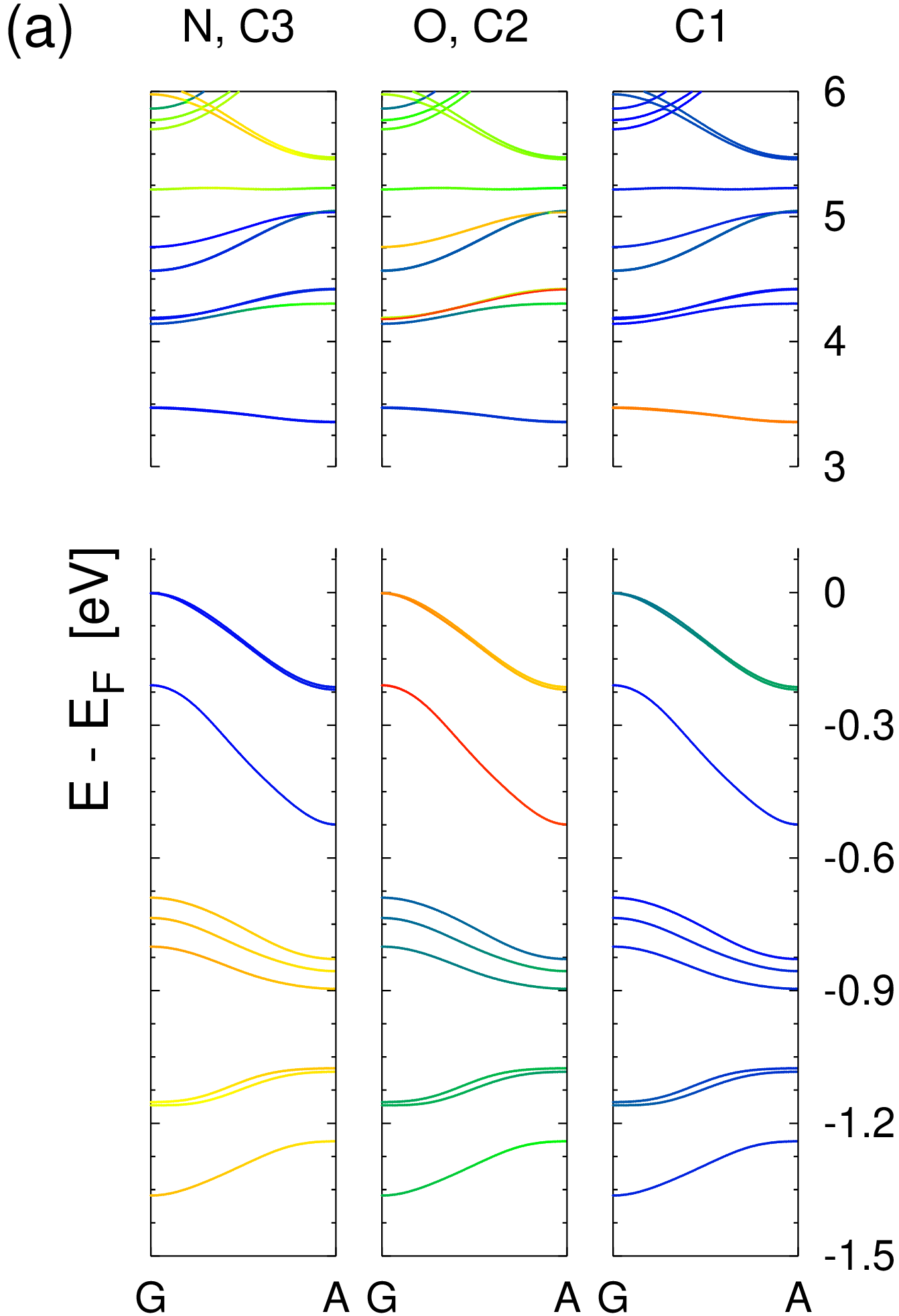} \hspace{0.2cm}
\includegraphics[scale=0.25,angle=-90.0]{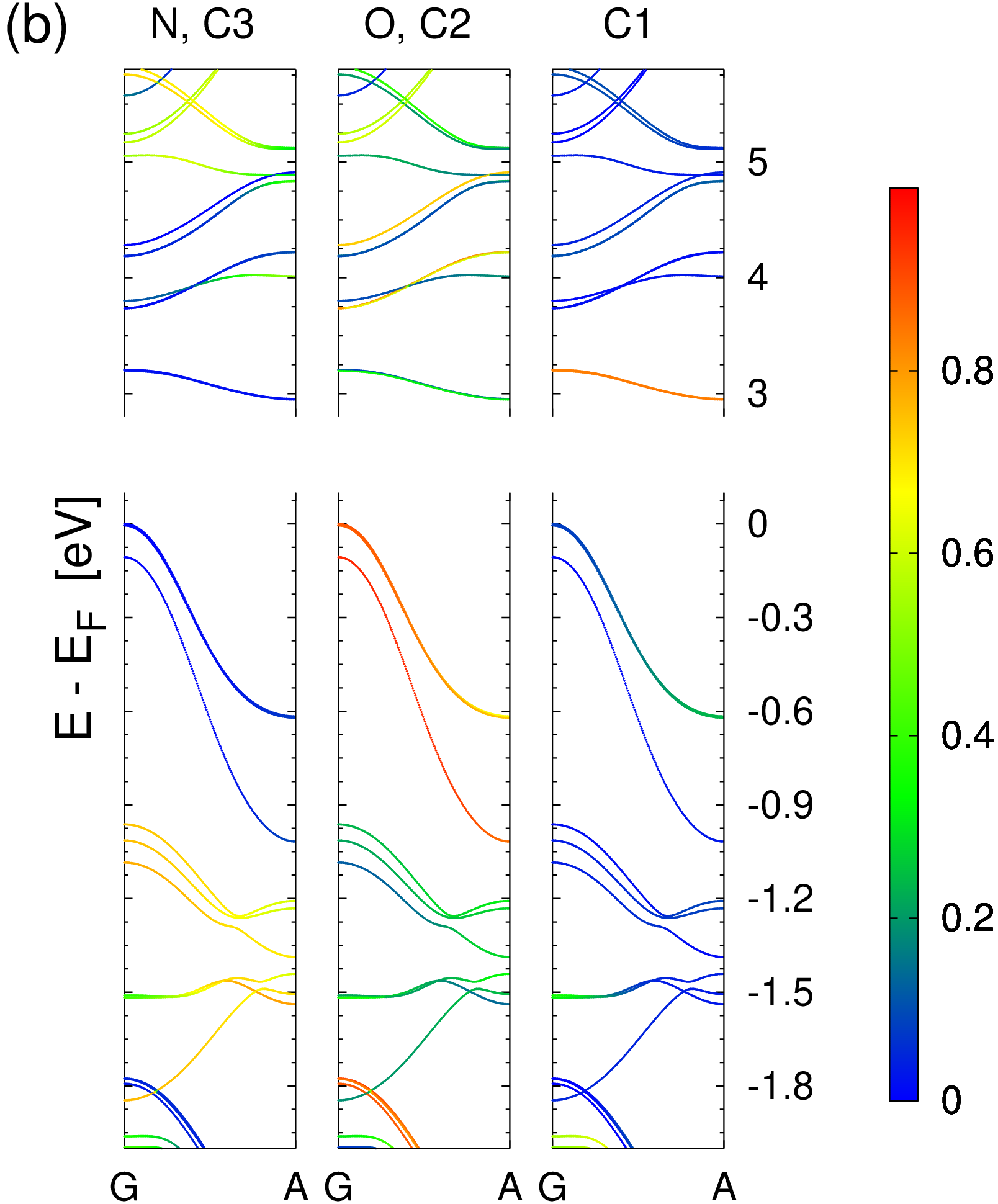}}
\caption{Band structure, projected at the  groups of atoms indexed in Fig. 1(c),
of the columnar structure for two intermolecular distances:
(a) 5.2 $\AA$ optimized for the GGA method, and (b) 4.6 $\AA$ optimized 
for the LDA method. The color scale represents a sum of the coefficients 
of the chosen Wannier functions in the wavefunction expansion.}
\label{b4}
\end{figure*}

\begin{figure}
\vspace{3cm}
\hspace{-2.5cm}  \includegraphics[scale=0.25,angle=-90.0]{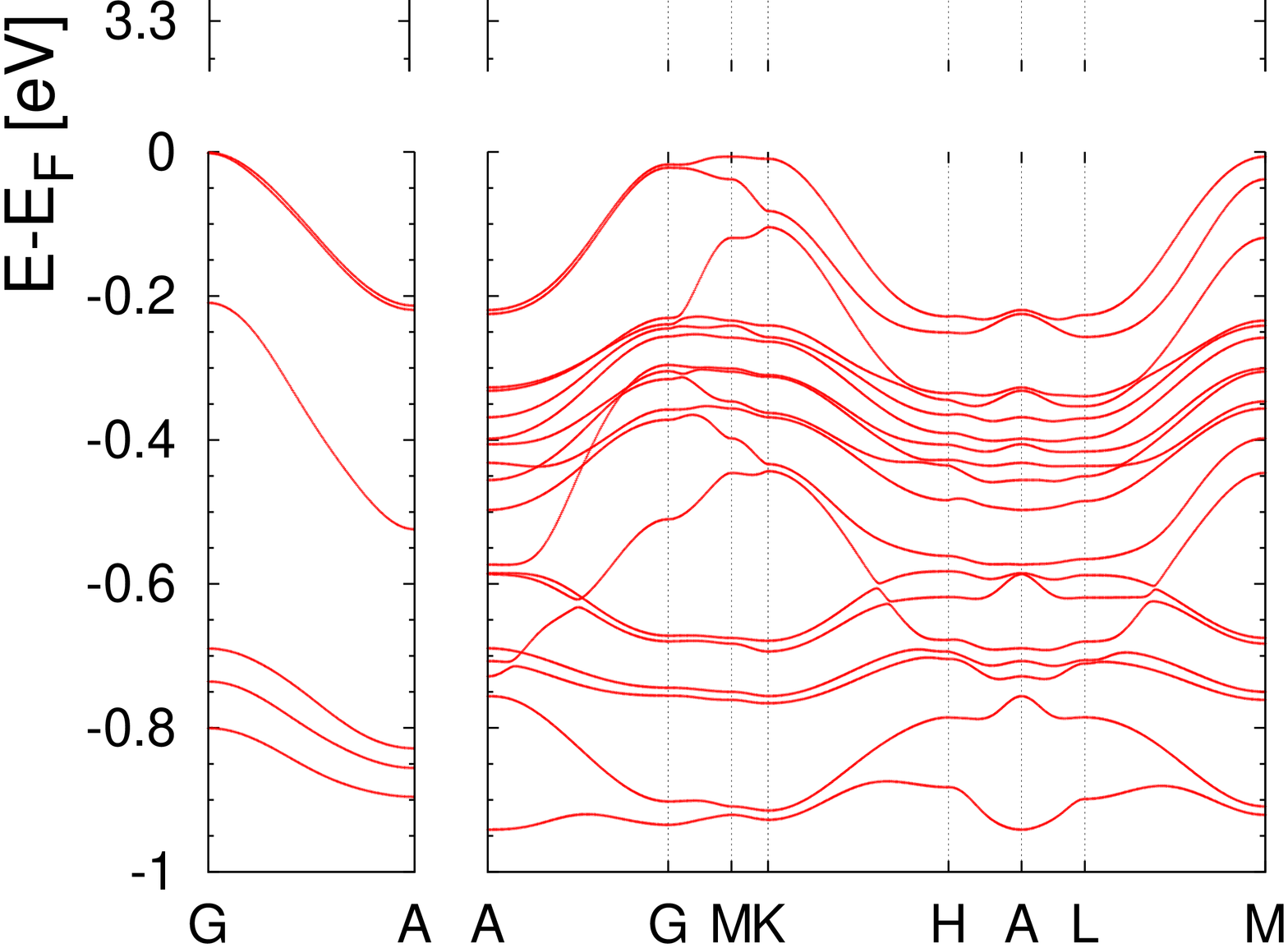}
\caption{Band structure of: (a) single molecular layer, (b) bulk structure of
molecular layers, (c) molecular wire.}
\label{b5}
\end{figure}

\begin{figure*}
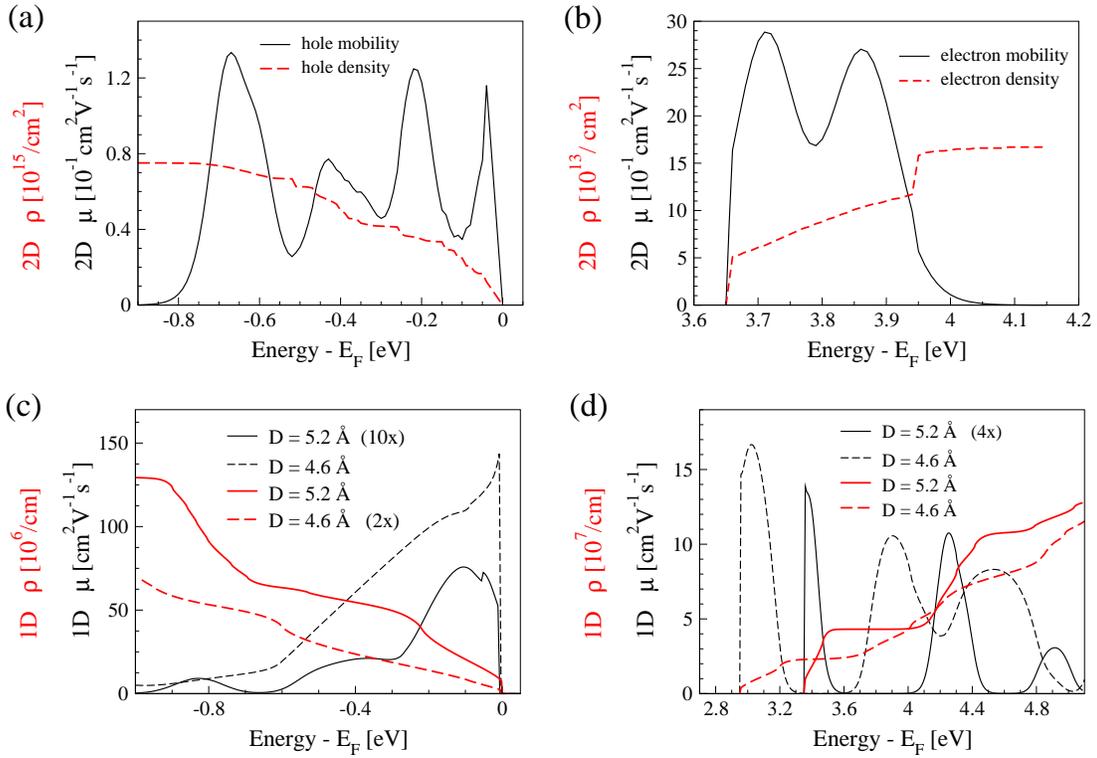

\centerline{
\includegraphics[scale=0.24,angle=0.0]{F6a.eps}\hspace{4mm}
\includegraphics[scale=0.24,angle=0.0]{F6b.eps}} \vspace{3mm}
\centerline{
\includegraphics[scale=0.24,angle=0.0]{F6c.eps} \hspace{4mm}
\includegraphics[scale=0.24,angle=0.0]{F6d.eps} }
\caption{2D carrier mobilities at 300 K and $\tau$=10 fs and carrier densities 
for (a) holes and (b) electrons, as well as 1D carrier mobilities and densities
for (c) holes and (d) electrons.
For better visibility, we scaled some curves as denoted in legends.}  
\label{b6}
\end{figure*}

\section{Results and discussion} 

Observation of the charge carrier paths, for the electrons and holes 
separately, sets a challenge for the experimental techniques. Hence, only 
a hypothesis has been derived from the experimental work on the reduced 
bi-molecular charge recombination in the metal-halide perovskite system.\cite{Snaith} 

In contrast, the theoretical methods allow to picture the carrier density 
in the real space by projection on the chosen atomic orbitals. 
Several molecular layers in the AA-type stacking pattern (one exactly on top
of the other) absorb light better than the single one, which 
would be transparent and photovoltaicly inefficient.
Additionally, the electronic transport across the films of the certain 
thickness might be affected by the stack--termination effects. For the polar 
layers, terminated with the nitrogen atoms on the top and mostly hydrogens at 
the bottom, one could expect different density of states for the outermost 
and inner layers. Surprisingly, the analysis of the projected DOS (PDOS) for 
the six layers system leads to the conclusion, that the hole and electron 
properties do not differ depending on the layer position in the stack. 

Fig. 2 shows the PDOS for the top layer only. The corresponding pictures for 
all the following layers do not differ in shape one from another.
Although, as described in our previous work they are shifted 
in the energy, according to the Stark effect.\cite{previous} 
The PDOS for all six layers is presented in Fig. S1 in the supporting 
information. The PDOS analysis indicates, that the electrons and 
holes localize at different parts of the building molecules. The electrons 
tend to localize at the central ring, while holes at the dipole terminal 
groups. This space separation is strongly pronounced for the electrons and 
light holes, and to some extend for the heavier holes.   
Similar space separation has been reported for the covalent organic frameworks
where two different molecules play the role of acceptor and donor.\cite{IHJ}

We model the effect of the electrodes on the terminal layers of the optically  
active material with the graphene sheets. This material is a good 
candidate for both the cathode and anode, due to its high polarizability 
and induced change of the work function.\cite{previous} The hydrogen bonds 
within the 2D networks are weak, and in fact, there are also possibilities 
to pattern at graphene without any intermolecular bonding.\cite{graphorder,graf2} 
Hence, we consider the separated molecular stacks with three 
different arrangements between the molecules and graphene. The top and side 
views are presented in Figs. 3(a)-(d). By breaking the planar connections, 
we gain the additional path for the holes, which now can move between the 
neighboring carboxy groups. As seen in Fig. 3(e), 
the application of the graphene electrodes do not affect the electron--hole 
path separation. The effect of molecular orientation with respect to the 
graphene axes, as for the cases presented in Figs. 3(a)-(c), turned out to 
be negligible. The PDOS for all other considered orientations 
is presented in Fig. S2 in the supporting information.

Breaking the planar bonds, as well as addition of the graphene 
leads, modifies the valence band character. 
Comparison of Fig. 2 and Fig. 3(e), shows that the top of the valence band
in the 2D case (Fig. 2) is built of the states localized at nitrogens and oxygens
- which means that the NCCH2- and COOH-group contribute to the states
at the same energy below the Fermi level. In contrast, in the 1D case (Fig. 3(e)),
the PDOS is built of the oxygen states close to the Fermi level and energetically 
deeper (around 2-3 eV below the valence band top) nitrogen states.   
This is a fingerprint of the energetic separation of the light and heavier hole. 
The same property is exhibited by the infinite 1D molecular stack, for 
which the band structures projected at the near atomic--centred Wannier 
functions are shown in Figs. 4(a) and (b) for two lattice constants.
The projection onto oxygens (the red color in the middle panel)   
shows that the contribution of the COOH group is energetically located just below
the Fermi level. While the projection to nitrogens (the orange color in the
leftmost panel) displays the energetic location of the contribution of the NCCH2 group, 
which is well below that of the COOH moieties.

\begin{figure*}
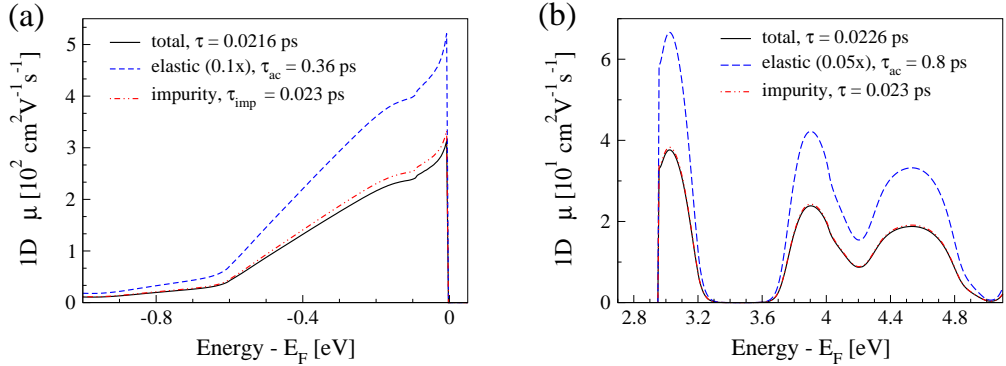

\centerline{
\includegraphics[scale=0.24,angle=0.0]{F7a.eps}\hspace{8mm}
\includegraphics[scale=0.24,angle=0.0]{F7b.eps}}
\caption{1D carrier mobilities, at the intermolecular distance 4.6 $\AA$,
for (a) holes and (b) electrons, calculated at 300 K
with the relaxation times obtained with the elastic scattering theory ($\tau_{ac}$),
the scattering on the ionic impurities ($\tau_{imp}$), and the both effects ($\tau$).
For better visibility, we scaled some curves as denoted in legends.}
\label{b7}
\end{figure*}

\begin{figure}
\includegraphics[scale=0.38,angle=0.0]{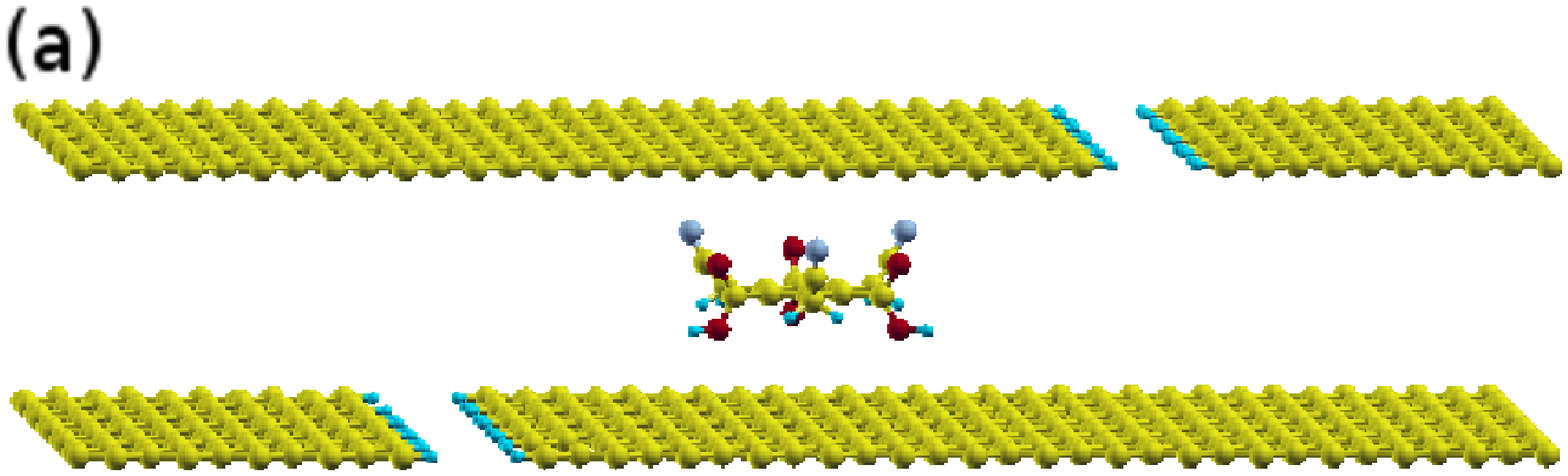}
\includegraphics[scale=0.33,angle=-90.0]{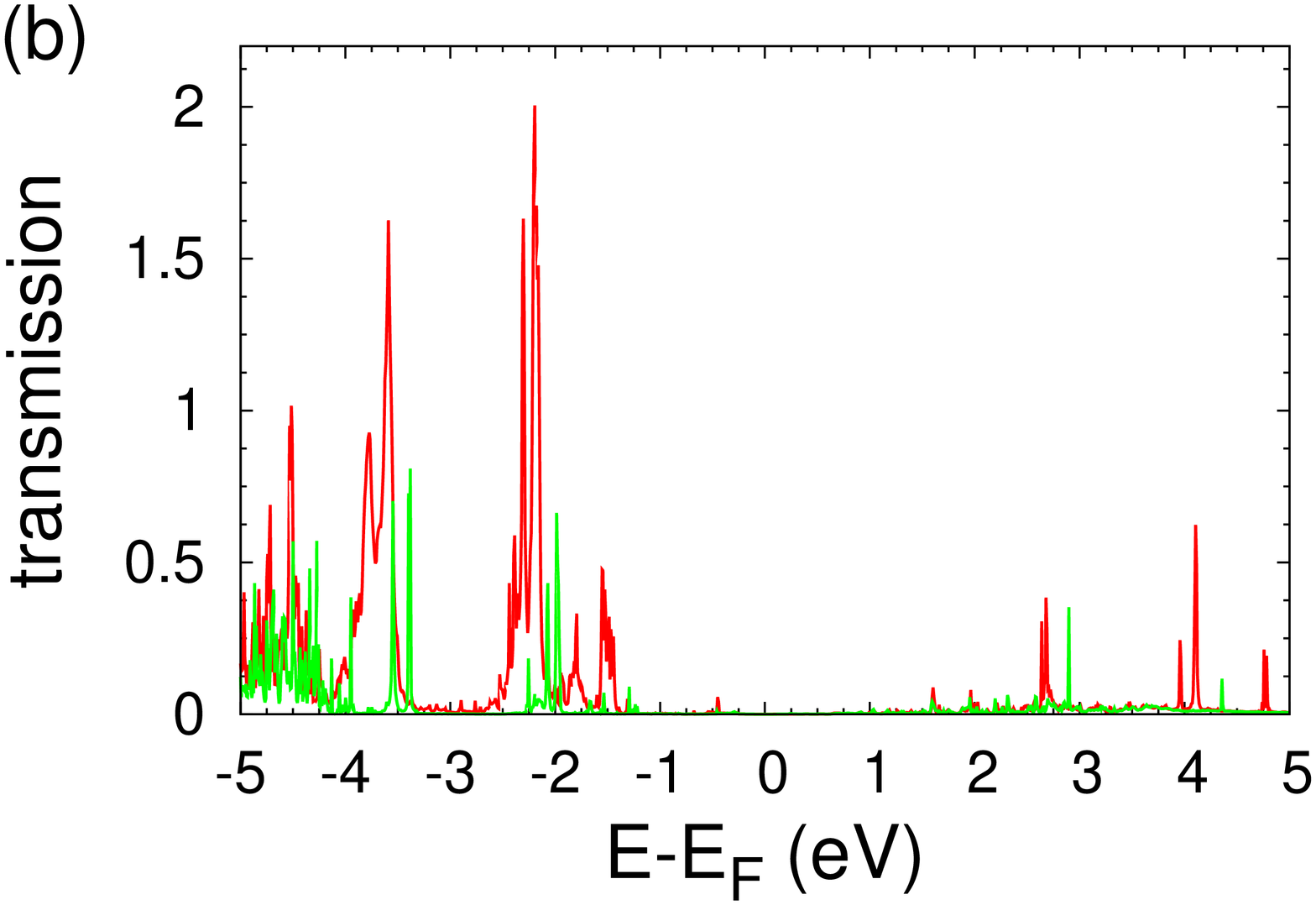}
\includegraphics[scale=0.33,angle=-90.0]{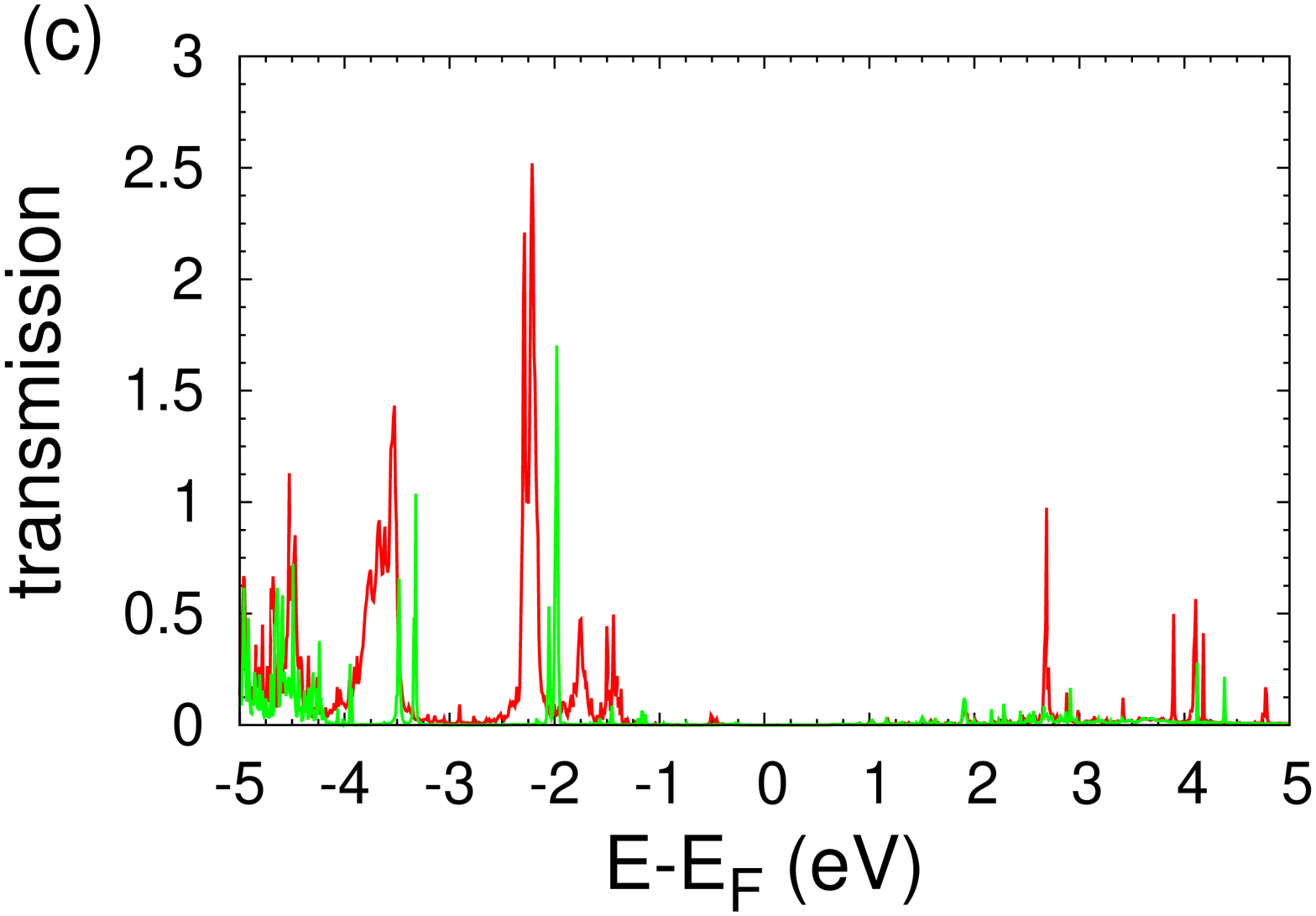}
\caption{Geometry of the simulated system with the graphene
leads (a), the transmission obtained with the GGA method (b), and with
the LDA method (c). Red (green) lines correspond to the stronger (weaker)
coupling between molecule and graphene sheets, i.e. 8.6 and 9.2 $\AA$,
repsectively.}
\label{b8}
\end{figure}

The bandwidth is a signature of the carrier mobility, and it is sensitive 
to the strength of the $p_z$-orbitals overlap. This can be 
seen from a comparison of the two above described band structures. 
The bandwidth of the light hole is about 0.5 eV, for the
larger separation between the $\pi$-rings (Fig. 4(a)), and about 1 eV, for the more
closely stacked case (Fig. 4(b)). The heavier hole extends to about 1.4 eV 
 or 1.85 eV for the GGA- or LDA-optimized ring separations, correspondingly.
These bandwidths are directly connected to the hole mobilities, which are 
the derivatives of the energy levels with respect to the lines in the k-space
(the change of the point in the Brillouin zone). In other words: fast carriers move  
along thick bands. The effect of the inter-ring separation for the electron mobilities
is less pronounced that that for the holes.

The energy barriers for the rotation of the dipole groups are expected to be small 
\-– especially at the elevated temperatures, when the devise is exposed to Sun shine. 
But in the technological reality, the columnar liquid crystals are densely packed, 
as well as the columns studied in this work are also placed closely - on the hydrogen 
bond distance in the 2D case. The 1D model is good enough for the analysis of the mobilities, 
but the working device will be more similar to the 2D case, where flipping 
the orientation of the dipole group is not as easy as in the 1D case.

The transport properties of the columnar and planar systems differ too.
Therefore, in Fig. 5, we compare the electronic structures of the 1D and 
2D systems, including also the model 3D bulk material, which connects 
the parameters of those two. For the efficient solar cell devices, it is 
desirable to achieve high conductivity across the layers, whereas insulating 
properties within the planes -- for the purpose of the energy 
dissipation reduction. Indeed, the band dispersions of the 2D system are 
very flat with respect to the 1D case. The band gap of the 2D molecular layer 
is very similar to the HOMO-LUMO gap of an isolated molecule, and decreases 
with the thickness of the film, to be the smallest for the infinite molecular 
wire.

The number of carriers (holes and electrons) per the energetic range decreases 
for the low dimensional systems. The mobilities are defined 
as the ratio of the conductivity, proportional to the band dispersion, and 
the carrier density. This is one of the reasons for the high 
carrier mobilities obtained in graphene, for example. Figs. 6(a)-(b) present 
the mobilities obtained for the 2D and Figs. 6(c)-(d) for the 1D molecular 
system, respectively.

The results for the 1D system are compared for the two intermolecular 
distances, see Figs. 6(c)-(d). As expected, the carrier mobilities 
for the 2D case (with molecules connected {\it via} the hydrogen bonds) are 
much lower than these for the 1D system (with the molecules in the 
$\pi$-stacking); even at larger (the GGA) lattice constant. This effect 
is enhanced for the holes, due to the vertical geometry of the dipole groups, 
which makes the intermolecular distances in the terminal parts smaller than 
in the central part. On the other hand, the inter--ring distances are equal 
to the lattice constant of the chain, thus much larger than for instance 
the interplanar distances in graphite ($\approx$3.4 $\AA$). Decreasing the 
lattice constant, the electron mobilities grow about 4 times, while the 
corresponding increase for the holes is about 15 times.  

Additionally, in the supporting information, we show the conductivities
scaled with the constant relaxation time and plotted for several
temperatures, for the 2D and 1D cases; 
see Figs. S4 and S5 in the supporting information, respectively.
For the 2D system, a decrease of the conductivity with the temperature is
much more strongly pronounced with respect to the 1D case. Moreover, we
observe a visible anisotropy of the hole conductivity along the zigzag
and armchair directions in our hexagonal molecular lattice. It is worth
to point out, that it is not the case for the electron mobilities.

In order to go beyond the mobilities parametrized with the relaxation time,
for the 1D infinite molecular column, we estimate the scattering effects 
according to the elastic and the ionic impurity contributions.  
We abandon the calculation of the optical-phonons contribution, because
in the organic systems this kind of scattering is much lower than that 
caused by the acoustic phonons.\cite{elast,napht,srep,Casian}
Taking into account only the elastic scattering, the calculated relaxation time for 
the molecular pillar with the intermolecular distance of 4.6 $\AA$ is 0.36 ps for holes 
and 0.8 ps for the electrons. These values of $\tau$ correspond to 
the moblities achieving 5000 cm$^{2}$V$^{-1}$s$^{-1}$
for the holes and 1360 cm$^{2}$V$^{-1}$s$^{-1}$ for the electrons.

In order to calculate the scattering from the ionic impurities,  
we used the formulae given in the methods section and assumed one ionic 
impurity with the charge 1 or -1 (which leads to the same result) per ten
molecules in the wire (n=0.1/(4.6$\AA$), $Z_{ion}$=1). The density
of free carriers is assumed 1 per the unit cell for the holes and electrons
($N_0$=1/(4.6$\AA$)). 
The relative dielectric constant $\varepsilon_r$ of our wire was 
found 1.2 \-- accoring to the static component of the real part of the dielectric
function.\cite{eps}       
The calculated relaxation time for the ionic impurity scattering only is 
23 fs for the holes and the electrons. 
Thus, the maximal values of the mobilities
achieve 300 cm$^{2}$V$^{-1}$s$^{-1}$ and 39 cm$^{2}$V$^{-1}$s$^{-1}$ for 
the holes and electrons, respectively. 

The ionic scattering takes over the acoustic contribution. Hence, the total
relaxation time is 21.6 fs for holes and 22.6 fs for the electrons. 
The mobilities rescalled according to the obtained relaxation time are
printed for holes and electrons in Figs. 7(a) and 7(b), respectively.    
Concluding, the elastic effects in our molecular wires are compared to these
in the organometal halide perovskites.\cite{srep} The addition
of the ionic defects \-- which are often in the molecular 
crystals\cite{elast,napht,penta} --  drastically increases the scattering 
and decreases the mobilities. Our analysis shows that, despite the presence 
of the ionic impurities, the mobilities in our 1D system are high in comparison 
to other organic systems. The previous works show that the mobilities in some 
organic materials excess 100 cm$^{2}$V$^{-1}$s$^{-1}$ 
at low temperatures.\cite{mob1,mob2} 
We underline, that the systems studied in this work are highly
ordered, which is another reason for the low scattering and it is known  
that the mobilities in organic crystals are very anisotropic.\cite{anisotropy}

We compare the mobilities for our systems with these of 
the organometal halide perovskites.\cite{Sanvito,srep,Snaith}  
The relaxtion time calculated for our 1D system leads to the mobilities 
\-- 300 and 39 cm$^{2}$V$^{-1}$s$^{-1}$ for the holes and electrons, respectively 
\-- and these values are larger than for the perovskites. 
For instance, for CH$_3$NH$_3$PbI3 in the cubic structure,
reported in Fig. 4 in Ref. [54], the maximal values for the holes achieve
12 cm$^{2}$V$^{-1}$s$^{-1}$ and 8 cm$^{2}$V$^{-1}$s$^{-1}$ for the electrons
\-- with the assumed relaxation time of 1 ps for the both types of charges.
These values agree with the measured mobilities.\cite{Snaith} 
However, the other {\it ab initio} calculations with the inclusion 
of the spin-orbit coupling, reported in Table 1 in Ref. [38], 
lead to largely overestimated values of 1400-2200 cm$^{2}$V$^{-1}$s$^{-1}$
for the holes and 570-800 cm$^{2}$V$^{-1}$s$^{-1}$ for the electrons.
The above mobilities were obtained with the same methods which were applied
in this study. The relaxation times obtained by Zhao et al.\cite{srep} 
are an order of magnitude larger than our values reported in Figs. 7(a) and 7(b). 
This overestimation results probably from large increase of the bandwidths 
when the spin-orbit coupling is included and 
the quasiparticle corrections not.\cite{picozzi}
 
A single molecule sandwiched between the terminated,
H-passivated graphene planes is shown in Fig. 8(a). 
This is the minimal model which represents the $\pi$-stacking 
order between the optical spacer (molecules) and the electrodes.
The zero-bias 
transmission function calculations are performed within the GGA and LDA 
schemes -- see Figs. 8(b) and (c), respectively. The two coupling strengths, 
related to the distances between the molecule and the planes, are considered. 
The stronger (weaker) coupling corresponds to the case, where the distances 
are set to 4.6 and 4.0 $\AA$ (5.2 $\AA$ and 4.6 $\AA$) for the upper and lower 
graphene-molecule separations, respectively. 
The value of the width of the band gap visible in 
the transmission function stays in a good agreement with the gap size in the 
PDOS obtained for the molecular stack, see Fig. 3(e). The coupling strength 
determines the widths of the transmission peaks, i.e. increasing the 
coupling broadens the peaks. 
These facts correspond to the discussed mobility features,
as well as the band structure derived conclusions. 
For the weaker coupling cases the positions of the peaks are slightly shifted  
towards the Fermi energy. This effect is related to the renormalization of the
energy levels of the system. 
The threshold bias for the 'dark current' is about 3 eV for the holes, 
while for the electrons it is approximately 2 times higher.

\section{Conclusions}

We found and discuss the unique transport properties in the $\pi$-stacked
aromatic-rings with the terminal groups possessing the dipole moment -- the 
separate paths for the electrons and holes. This feature could be a solution 
for the charge recombination problem in the solar cell devices. The electronic 
diffusion paths cut through the aromatic rings, whereas the holes hop between 
the neighboring dipole groups. We verified theoretically, that connecting 
the system to the graphene electrodes does not destroy the above separation.  
The ferroelectric and path-separation properties are not necessarily directly 
connected to each other. The driving force for the electron-hole separation is 
not the dipole moment itself, but the excitonic character of the system - 
with the donor states at the $p$-electron rich atoms and the acceptor states at 
the aromatic ring. 

The proposed materials can be realized in 2D or 1D, as the layers or molecular 
wires, correspondingly. For the simplicity, we use the smallest model: 
one benzene ring, and the cyano groups for the ferroelectric properties,
and the carboxy groups for the formation of the hydrogen bonds within the 
planes. This choice of the planar bonds is dictated by the requirement of 
a high planar resistance. On the other hand, the carrier mobilities, for 
the electrons and holes across the layers, respectively, are one and two 
orders of magnitude larger than the planar ones.

In comparison with the organometal halide perovskites, 
reported in Ref. [54], our mobilities along the $\pi$-stacked 
1D columns are an order of magnitude larger for holes and about 5 times 
larger for electrons. These values are obtained in the presence 
of one charged impurity per ten molecules, which gives large contribution 
to the scattering mechanisms.
 \\[0.5cm]

{\bf Supporting Information Description} \\

Further details of the calculations, (S1) DOS projected at groups of atoms indexed in Fig. 1(c),
for each of six monolayers in the AA-type stacking without the leads,
(S2) the same for four monolayers between the graphene leads, (S3) comparison of that for
the first molecular layer in the three stackings: AA-type, R30-deg-type and AB-type,
(S4) conductivity function plots in the 2D case, (S5) the same for the 1D case.
\\[0.5cm]

{\bf Acknowledgements} \\

This work has been supported by The National Science Centre of Poland
(the Projects No. 2013/11/B/ST3/04041 and DEC-2012/07/B/ST3/03412).
Calculations have been performed in the Cyfronet Computer Centre using 
Prometheus computer which is a part of the PL-Grid Infrastructure, 
and by part in the Interdisciplinary Centre of
Mathematical and Computer Modeling (ICM). \\[0.5cm]


\begin{thebibliography}{99}

\bibitem{pedot:pss}
Rutledge S. A.; Helmy, A. S.
 Carrier Mobility Enhancement in Poly(3,4-ethylenedioxythiophene)-poly
(styrenesulfonate) Having Undergone Rapid Thermal Annealing.
{\it J. App. Phys.} {\bf 2013}, {\it 114}, 133708.

\bibitem{GON}
Liu, J.; Kim, G. H.; Xue, Y.; Kim, J. Y.; Baek, J. B.; Durstock, M.; Dai, L..
Graphene Oxide Nanoribbon as Hole
Extraction Layer to Enhance Efficiency and Stability of Polymer Solar Cells.
{\it Adv. Mater.} {\bf 2014}, {\it 26}, 786-790.

\bibitem{Olga}
Malinkiewicz, O.; Rold\'an-Carmona, C.; Soriano, A.; Bandiello, E.; Camacho, L.;
Nazeeruddin, M. K.; Bolink, H. J.
Metal-Oxide-Free Methylammonium Lead Iodide Perovskite-Based Solar Cells: 
the Influence of Organic Charge Transport Layers.
{\it Adv. En. Mater.} {\bf 2014}, {\it 4}, 1400345.

\bibitem{cpl1}
Shkrob, I. A.; Marin, T. W.
Charge Trapping in Photovoltaically Active Perovskites and Related
Halogenoplumbate Compounds.
{\it Chem. Phys. Lett.} {\bf 2014}, {\it 5}, 1066-1071. 

\bibitem{jpcl2}
Tachikawa, T.; Karimata, I.; Kobori, Y.
Surface Charge Trapping in Organolead Halide Perovskites Explored
by Single-Particle Photoluminescence Imaging.
{\it J. Phys. Chem. Lett.} {\bf 2015}, {\it 6}, 3195-3201.

\bibitem{natnano1}
Stranks S. D.; Snaith, H. J. 
Metal-Halide Perovskites for Photovoltaic and Light-Emitting Devices.
{\it Nature Nanotech.} {\bf 2015}, {\it 10}, 391-402.

\bibitem{acsnano1}
Leijtens, T.; Stranks, S. D.; Eperon, G. E.; Lindblad, R.; Johansson, E. M. J.;
McPherson, I. J.; Rensmo, H.; Ball, J. M.; Lee, M. M.; Snaith, H. J.
Electronic Properties of Meso-Superstructured and Planar
Organometal Halide Perovskite Films: Charge Trapping, Photodoping, and
Carrier Mobility.
{\it ACS Nano} {\bf 2014}, {\it 8}, 7147-7155.

\bibitem{pccp1}
Wang, Y.; Wang, H.-Y.; Yu, M.; Fu, L.-M.; Qin, Y.; Zhang, J.-P.; Ai, X.-C.
Trap-Limited Charge Recombination in Intrinsic Perovskite Film and
Meso-Superstructured Perovskite Solar Cells and The Passivation Effect of
The Hole-Transport Material on Trap States.
{\it Phys. Chem. Chem. Phys.} {\bf 2015}, {\it 17}, 29501-29506.

\bibitem{model1}
Stranks, S. D.; Burlakov, V. M.; Leijtens, T.; Ball, J. M.; Goriely, A.; Snaith, H. J.
Recombination Kinetics in Organic-Inorganic Perovskites: Excitons, Free Charge,
and Subgap States.
{\it Phys. Rev. Appl.} {\bf 2014}, {\it 2}, 034007-8.

\bibitem{science1}
Shi, D.; Adinolfi, V.; Comin, R.; Yuan, M.; Alarousu, E.;
Buin, A.; Chen, Y.; Hoogland, S.; Rothenberger, A.;
Katsiev, K. et al.
Low Trap-State Density and Long Carrier Diffusion in Organolead
Trihalide Perovskite Single Crystals.
{\it Science} {\bf 2015}, {\it 347}, 519-522.

\bibitem{AFM1}
Milot, R. L.; Eperon, G. E.; Snaith, H. J.; Johnston, M. B.; Herz, L. M.
Temperature-Dependent Charge-Carrier Dynamics in
CH3NH3PbI3 Perovskite Thin Films.
{\it Adv. Funct. Mater.} {\bf 2015}, {\it 25}, 6218-6227.

\bibitem{bulkhetero}
Xiao, Z.; Yuan, Y.; Yang, B.; VanDerslice, Y.; Chen, J.; Dyck, O.;
Duscher, G.; Huang, J. 
Universal Formation of Compositionally Graded Bulk
Heterojunction for Efficiency Enhancement in Organic Photovoltaics.
{\it Adv. Mater.} {\bf 2014}, {\it 26}, 3068-3075.

\bibitem{IHJ}
Calik, M.; Auras, F.; Salonen, L. M.; Bader, K.; Grill, I.; Handloser, M.;
Medina, D. D.; Dogru, M.; L\"obermann, F.; Trauner, D. et al.
Extraction of Photogenerated Electrons and Holes from a Covalent
Organic Framework Integrated Heterojunction.
{\it J. Am. Chem. Soc.} {\bf 2014}, {\it 136}, 17802-17807.

\bibitem{Bein-3}
Wuttke, S.; Dietl, C.; Hinterholzinger, F. M.; Hintz, H.;
Langhals, H.; Bein, T.
Turn-On Fluorescence Triggered by Selective Internal Dye Replacement in MOFs.
{\it Chem. Commun.} {\bf 2014}, {\it 50}, 3599-3601.

\bibitem{Bein-4}
Dogru, M.; Handloser, M.; Auras, F.; Kunz, T.; Medina, D.;
Hartschuh, A.; Knochel, P.; Bein, T.
A Photoconductive Thienothiophene-Based Covalent Organic
Framework Showing Charge Transfer Towards Included Fullerene.
{\it Angew. Chem. Int. Ed.} {\bf 2013}, {\it 52}, 2920-2924.

\bibitem{Bein-5}
Dogru, M.; Bein, T.
On The Road Towards Electroactive Covalent Organic Frameworks.
{\it Chem. Commun.} {\bf 2014}, {\it 50}, 5531-5546.

\bibitem{cascade}
Tan, Z.-k.; Johnson, K.; Vaynzof, Y.; Bakulin, A. A.; Chua, L.-L.;
Ho, P. K. H.; Friend, R. K. 
Suppressing Recombination in Polymer Photovoltaic
Devices via Energy-Level Cascades.
{\it Adv. Mater.} {\bf 2013}, {\it 25}, 4131-4138.

\bibitem{workOH}
Khazaei, M.; Arai, M.; Sasaki, T.; Ranjbar, A.; Liang, Y.; Yunoki, S.
OH-Terminated Two-Dimensional Transition Metal Carbides and Nitrides
as Ultralow Work Function Materials.
{\it Phys. Rev. B} {\bf 2015}, {\it 92}, 075411-10.

\bibitem{previous}
Wierzbowska, M.; Wawrzyniak-Adamczewska, M.
Cascade Donor-Acceptor Organic Ferroelectric Layers, Between Graphene Sheets, 
for Solar Cell Applications.
eprint {\bf 2015}, arXiv:cond-mat:1510.05220.

\bibitem{LC}
Sergeyev, S.; Pisula, W.; Geerts,  Y. H.
Discotic Liquid Crystals: A New Generation of Organic Semiconductors.
{\it Chem. Soc. Rev.} {\bf 2007}, {\it 36}, 1902-1929. 

\bibitem{graphorder}
Reinert, F.
Graphene Gets Molecules Into Order.
{\it Nature Phys.} {\bf 2013}, {\it 9}, 321-322. 

\bibitem{Lublin}
Zhang, X.; Li, N.; Gu, G.-C.; Wang, H.; Nieckarz, D.; Szabelski, P.; He, Y.; Wang, Y.; 
Xie, C.; Shen, Z.-Y. et al.
Controlling Molecular Growth between Fractals and Crystals on Surfaces.
{\it ACS Nano} {\bf 2015}, {\it 9}, 11909-11915. 

\bibitem{Bein-1}
Medina, D. D.; Rotter, J. M.; Hu, Y.; Dogru, M.; Werner, V.; Auras, F.
J. T. Markiewicz, P. Knochel, T. Bein,
Room Temperature Synthesis of Covalent–Organic Framework Films
through Vapor-Assisted Conversion.
{\it J. Am. Chem. Soc.} {\bf 2015}, {\bf 137}, 1016-1019.

\bibitem{Berkeley}
Zhang, K.-D.; Tian, J.; Hanifi, D.; Zhang, Y.; Sue, A. C.-H.; Zhou, T.-Y.; Zhang, L.;
Zhao, X.; Liu, Y.; Li, Z.-T.
Toward a Single-Layer Two-Dimensional Honeycomb Supramolecular Organic
Framework in Water.
{\it J. Am. Chem. Soc.} {\bf 2013}, {\it 135}, 17913-17918. 

\bibitem{Horiuchi}
Horiuchi, S.; Tokura, Y.
Organic Ferroelectrics.
{\it Nature Mater. (Review)}, {\bf 2008}, {\it 7}, 357-366.

\bibitem{Sobol}
Sobolewski, A. L.
Organic Photovoltaics Without p–n Junctions: A Computational Study of Ferroelectric Columnar
Molecular Clusters.
{\it Phys. Chem. Chem. Phys.} {\bf 2015}, {\it 17}, 20580-8.

\bibitem{Horiuchi-2}
Horiuchi, S.; Kumai, R.; Tokura, Y.
High-Temperature and Pressure-Induced Ferroelectricity in Hydrogen-Bonded
Supramolecular Crystals of Anilic Acids and 2,3-Di(2-pyridinyl)pyrazine.
{\it J. Am. Chem. Soc.} {\bf 2013}, {\it 135}, 4492-4500.

\bibitem{Horiuchi-3}
Horiuchi, S.; Kobayashi, K.; Kumai, R.; Ishibashi, S.
Ionic Versus Electronic Ferroelectricity on Donor-Acceptor Molecular Sequence.
{\it Chem. Lett. (Highlight Review)} {\bf 2014}, {\it 43}, 26-35.

\bibitem{qe}
Giannozzi, P.; Baroni, S.; Bonini, N.; Calandra, M.; Car, R.; Cavvazzoni, C.;  
Ceresoli, D.; Chiarotti, G. L.; Cococcioni, M.; Dabo, I.
QUANTUM ESPRESSO: A Modular and Open-Source Software Project for
Quantum Simulations of Materials.
{\it J. Phys. Condens. Matter} {\bf 2009}, {\it 21}, 395502-19.

\bibitem{PBE}
Perdew, J. P.; Burke, K.; Ernzerhof, M.
Generalized Gradient Approximation Made Simple.
{\it Phys. Rev. Lett.} {\bf 1997}, {\it 78}, 1396.

\bibitem{w90}
Mostofi, A. A.; Yates, Y. R.; Lee, Y. S.; Souza, I.; Vanderbilt, D.; Marzari, N.
wannier90: A Tool for Obtaining Maximally-Localised Wannier Functions.
{\it Comput. Phys. Commun.} {\bf 2008}, {\it 178}, 685-699.

\bibitem{wan}
Marzari, N.; Vanderbilt, D.
Maximally Localized Generalized Wannier Functions for Composite Energy Bands.
{it Phys. Rev. B} {\bf 1997}, {\it 56}, 12847.

\bibitem{RMP}
Marzari, N.; Mostofi, A. A.; Yates, J. R.; Souza, I.; Vanderbilt, D.
Maximally Localized Wannier Functions: Theory and Applications.
{\it Rev. Mod. Phys.} {\bf 2012}, {\it 84}, 1419-1475.

\bibitem{boltz}
Pizzi, G.; Volja, D.; Kozinsky, B.; Fornari, M.; Marzari, N.
BoltzWann: A Code for The Evaluation of Thermoelectric and Electronic Transport
Properties with a Maximally-Localized Wannier Functions Basis.
{\it Comp. Phys. Comm.} {\bf 2014}, {\it 185}, 422-429.

\bibitem{bardeen}
Bardeen, J.; Shockley, W.
Deformation Potentials and Mobilities in Non-Polar Crystals.
{\it Phys. Rev.} {\bf 1950}, {\it 80}, 72-80.

\bibitem{elast}
Tang, L.; Long, M.-Q.; Wang, D.; Shuai, Z.-G.
The Role of Acoustic Phonon Scattering in Charge Transport in
Organic Semiconductors: A First-Principles Deformation-Potential Study.
{\it Sci. China Ser. B-Chem.} {\bf 2009}, {\it 52}, 1646-1652. 

\bibitem{napht} 
Wang, L. J; Peng, Q.; Li, Q. K.; Shuai, Z. G. 
Roles of Inter- and Intramolecular Vibrations and Band-Hopping Crossover in
The Charge Transport in Naphthalene Crystal. 
{\it J. Chem. Phys.} {\bf 2007}, {\it 127}, 044506. 

\bibitem{srep}
Zhao, T.; Shi, W.; Xi, J.; Wang, D.; Shuai, Z.
Intrinsic and Extrinsic Charge Transport in CH3NH3PbI3 Perovskites Predicted from
First-Principles.
{\it Scientific Reports} {\bf 2015}, {\it 6}, 19968.

\bibitem{Casian}
Casian, A.; Dusciac, V.; Coropceanu, Iu.
Huge Carrier Mobilities Expected in Quasi-One-Dimensional Organic Crystals.
{\it Phys. Rev. B} {\bf 2002}, {\it 66}, 165404.

\bibitem{penta}
Jurchescu, O. D.; Baas, J.; Palstra, T. T. M.
Effect of Impurities on The Mobility of Single Crystal Pentacene. 
{\it Appl. Phys. Lett.} {\bf 2004}, {\it 84}, 3061-3063.

\bibitem{ion}
Chattopadhyay, D.; Queisser, H. J. 
Electron-Scattering by Ionized Impurities in Semiconductors. 
{\it Rev. Mod. Phys.} {\bf 1981}, {\it 53}, 745-768.

\bibitem{eps}
QuantumESPRESSO. www.quantum-espresso.org (accessed March 10, 2016).

\bibitem{siesta1}
Ordej{\'o}n, P.; Artacho, E.; Soler, J. M.
Self-Consistent Order-N Density-Functional Calculations for Very Large Systems.
{\it Phys. Rev. B}, {\bf 1996}, {\it 53}, R10441.

\bibitem{siesta2}
Soler, J. M.; Artacho, E.; Gale, J. D.; Garc\'{\i}a, A.; Junquera, J.; Ordej{\'o}n, P.;
S{\'a}nchez-Portal, D.
The SIESTA Method for Ab Initio Order-N Materials Simulation.
{\it J. Phys. Condens. Matter} {\bf 2002}, {\it 14}, 2745.

\bibitem{Brandbyge02} 
Brandbyge, M.; Mozos, J. L.; Ordej\'on, P.; Taylor, J.; Stokbro, K.
Density-Functional Method for Nonequilibrium Electron Transport.
{\it Phys. Rev. B} {\bf 2002}, {\it 65}, 165401.

\bibitem{KS1}
Hohenberg, P.; Kohn, W. 
Inhomogeneous Electron Gas.
{\it Phys.~Rev.}, {\bf 1964}, {\it 136}, {B864}.

\bibitem{KS2}
Kohn, W.; Sham, L. J. 
Self-Consistent Equations Including Exchange and Correlation Effects.
{\it Phys.~Rev.} {\bf 1965}, {\it 140}, A1133.

\bibitem{CA} 
Ceperley, D. M.; Alder, B. J.
Ground State of the Electron Gas by a Stochastic Method.
{\it Phys. Rev. Lett.} {\bf 1980}, {\it 45}, 566.

\bibitem{Snaith}
Wehrenfennig, C.; Eperon, G. E.; Johnston, M. B.; Snaith, H. J.; Herz,  L. M.
High Charge Carrier Mobilities and Lifetimes in Organolead Trihalide Perovskites. 
{\it Adv. Mater.} {\bf 2014}, {\it 26}, 1584-1589.

\bibitem{graf2}
Colson, J. W.; Woll, A. R.; Mukherjee, A.; Levendorf, M. P.; Spitler, E. L.;
Shields, V. B.; Spencer, M. G.; Park, J.; Dichtel, W. R. 
Oriented 2D Covalent Organic Framework Thin Films on Single-Layer Graphen.
{\it Science} {\bf 2011}, {\it 332}, 228-231.

\bibitem{mob1}
Sakanoue, T.; Sirringhaus, H.
Band-Like Temperature Dependence of Mobility in a Solution-Processed
Organic Semiconductor.
{\it Nat. Mater.} {\bf 2010}, {\it 9}, 736-740.

\bibitem{mob2}
Karl, N.; Kraft, K.-H.; Marktanner, J.; M\"unch, M.; Schatz, F.; Stehle, R.; Uhde, H.-M.
Fast Electronic Transport in Organic Molecular Solids. 
{\it J. Vac. Sci. Technol. A} {\bf 1999}, {\it 17}, 2318-2328.

\bibitem{anisotropy}
de Wijs, G. A.; Mattheusa, C. C.; de Groot, R. A. Palstra, T. T. M.
Anisotropy of The Mobility of Pentacene from Frustration.
{\it Synth. Met.} {\bf 2003}, {\it 139}, 109-114.

\bibitem{Sanvito}
Motta, C.; Mellouhi, F. E.; Sanvito, S.
Charge Carrier Mobility in Hybrid Halide Perovskites.
{\it Sci. Rep.} {\bf 2015}, {\it 5}, 12746.

\bibitem{picozzi}
Bokdam, M.; Sander, T.; Stroppa, A.; Picozzi, S.; Sarma, D. D.; 
Franchini, C.; Kresse, G.
Role of Polar Phonons in the Photo Excited State of Metal Halide Perovskites.
eprint {\bf 2015}, arXiv:cond-mat/1512.05593. 

\end{thebibliography}
\end{document}